# PANDORA a New Facility for Interdisciplinary In-Plasma Physics

D. Mascali[1], A. Musumarra[1,2a], F. Leone[1,2,4], F.P. Romano[1,3], A. Galatá[5], S. Gammino[1], and C. Massimi[6,7]

[1] INFN-Laboratori Nazionali del Sud, I-95123 Catania, Italy
[2] Department of Physics and Astronomy University of Catania, I-95123 Catania, Italy
[3] CNR-IBAM, I-95124 Catania, Italy
[4] INAF-OACT, I-95123 Catania, Italy
[5] INFN-Laboratori Nazionali di Legnaro, Legnaro I-35020, Italy
[6] Department of Physics and Astronomy University of Bologna, I-40126 Bologna, Italy
[7] INFN-Bologna, I-40126 Bologna, Italy

**Abstract.** PANDORA, Plasmas for Astrophysics, Nuclear Decays Observation and Radiation for Archaeometry, is planned as a groundbreaking new facility based on a state-of-the-art plasma trap confining extremely energetic plasma for performing interdisciplinary research in the fields of Nuclear Astrophysics, Astrophysics, Plasma Physics and Applications in Material Science and Archaeometry: the plasmas become the environment for measuring, for the first time, nuclear decays rates in stellar-like condition (such as $^7$Be decay and beta-decay involved in s-process nucleosynthesis), especially as a function of the ionization state of the plasma ions. These studies are of paramount importance for addressing several astrophysical issues in both stellar and primordial nucleosynthesis environment (e.g. determination of solar neutrino flux and $^7$Li Cosmological Problem), moreover the confined energetic plasma will be a unique light source for high performance stellar spectroscopy measurements in the visible, UV and X-ray domains, offering advancements in observational astronomy. As to magnetic fields, the experimental validation of theoretical first and second order Landé factors will drive the layout of next generation polarimetric units for the high resolution spectrograph of the future giant telescopes. In PANDORA new plasma heating methods will be explored, that will push forward the ion beam output, in terms of extracted intensity and charge states. More, advanced and optimized injection methods of ions in an ECR plasma will be experimented, with the aim at optimizing its capture efficiency. This will be applied to the ECR-based Charge Breeding technique, that will improve the performances of the SPES ISOL-facility currently installed at Laboratori Nazionali di Legnaro-INFN. Finally, PANDORA will be suitable for energy conversion, making the plasma as an exceptional source of electromagnetic radiation, for applications in Material Science and Archaeometry.

**PACS.** 52.72.+v Laboratory studies of space- and astrophysical-plasma processes – 26. Nuclear Astrophysics – 23.40.-s Beta decay; Electron Capture – 95.55.Qf Photometric, polarimetric and spectroscopic instrumentation – 52.50.-b Plasma production and heating – 52.65.-y Plasma simulation – 52.70.-m Plasma diagnostics techniques and instrumentation

## 1 Introduction

Study and use of plasmas in science and industry is significantly growing. Although magnetically confined plasmas have been in the recent past investigated more for energy purposes, boosting the research in the field of high-density, high-temperature, long-living magnetoplasmas self sustaining nuclear fusion reactions, they still represent environments of relevant interest for interdisciplinary research in Astrophysics, Nuclear Astrophysics, etc. Compact magnetic traps have been mostly used as ion sources since 70s: such traps (called "B-minimum" machines, from the peculiar shape of the magnetic field) have supported the growing request of intense beams of multicharged ions coming from both fundamental science (nuclear and particle Physics especially) and applied research. Inside these machines[1–4] schematically depicted in figure 1 a dense and hot plasma, made of multicharged ions immersed in a dense cloud of energetic electrons, is confined by multi-Tesla magnetic fields and resonantly heated by some kWs of microwave power in the 2.45-28 GHz frequency range.

The plasmas reach $n_e \sim 10^{11}$-$10^{14}$ cm$^{-3}$, $T_e \sim 0.1$-$100$ keV of electron density and temperature, respectively.

These plasmas are Magneto Hydro Dynamically (MHD) stable [2], magnetized, living several hours or days with on average constant local density and temperature, so they represent an attractive environment for the studies envisaged in the framework of a new facility.

[a] corresponding author: musumarra@lns.infn.it

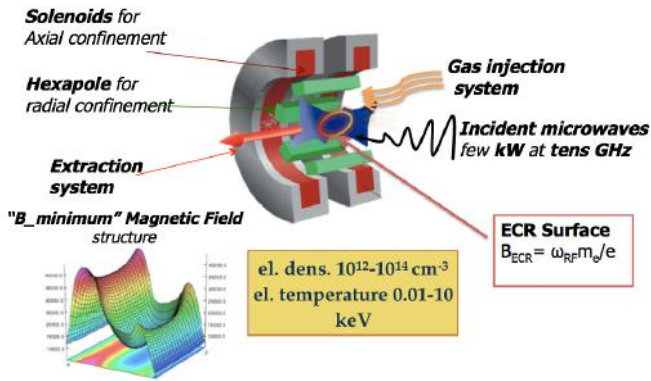

**Fig. 1.** Sketch of the experimental setup of a B-minimum magnetic trap, with typical values of density and temperature achievable in last-generation machines.

The goal of the PANDORA project (Plasmas for Astrophysics, Nuclear Decays Observation and Radiation for Archaeometry) is to establish a new experimental facility based on an innovative plasma trap confining energetic, dense/over-dense plasma. The plasma will be produced inside a compact magnetic trap by microwaves and microwave-driven plasma waves in the 18-28 GHz range, where several important physics cases can be afforded[3]:

i) **Nuclear Astrophysics**: plasma becomes the environment for measuring, for the first time, nuclear decays rates in stellar-like conditions (starting from $^7$Be decay and beta-decay involved in s-process nucleosynthesis), especially as a function of the in-plasma ionization state. These studies are of paramount importance for addressing several astrophysical issues in both stellar and primordial nucleosynthesis environment (e.g. determination of solar neutrino flux and $^7$Li Cosmological Problem).

ii) **Astrophysics**: the confined energetic plasma will be a unique and valuable light source complementing high performance stellar spectroscopy measurements for a better understanding of spectropolarimetric observations in the visible, UV and X-ray domains, offering advancements in observational astronomy. As to magnetic fields, the experimental validation of theoretical first and second order Landé factors will drive the layout of the high resolution spectropolarimeters of the future giant telescopes such as HIRES of the 39 m European-Extremely Large Telescope of the European Southern Observatory. Even more pioneering will be the study of the polarized X-ray emission, theoretically expected from compact and strongly magnetized stars.

iii) **Nuclear Physics**: new heating methods will be explored, that will push forward the ion beam output, in terms of extracted intensity and charge states. More, advanced and optimized injection methods of ions in an ECR plasma will be experimented, with the aim at optimizing its capture efficiency. This will be applied to the ECR-based charge breeding technique, that will improve the performances of an ISOL facility like the SPES facility now under commissioning at INFN-Laboratori Nazionali di Legnaro (Padova).

iv) **Intense x-ray source for Material Science**: the new plasma heating scheme is suitable for energy conversion making the plasma as an exceptional source of electromagnetic radiation, for applications in material science and Archaeometry.

## 2 ECRIS state-of-the-art

It is well-known that plasma physics finds its largest area of application in the controlled nuclear fusion field; on the other hand, compact plasma machines working as ion sources have played a crucial role in developing nuclear physics facilities. Among the various types of ion sources developed since 1950s, Electron Cyclotron Resonance Ion Sources (ECRIS) [4] are the most performing ones, supporting the growing request of intense beams of multi-charged ions coming from both fundamental science (nuclear and particle Physics especially [5]) and applied research (neutrons spallation sources [6], sub-critical nuclear reactors [7–9], hadrotherapy facilities [10], material treatments, ion implantation). The ECRIS development path has been based [11] on the mere magnetic field, microwave source frequency and power boosting, now deemed saturating. ECRIS of 3rd generation are capable to produce several tens of $\mu$A of Ar$^{14+}$ or Xe$^{34+}$ [12] and mA beams of others multiply-charged ions [13]; these beams can be extracted by plasmas reaching $n_e \sim 10^{13}$ cm$^{-3}$ of electron density. The density especially represents the most critical aspect. Its increase has required the pumping wave frequency to be increased as well, in order to move upward the critical density boundary. Large plasma densities are mandatory for producing intense currents of multi-charged ions. But ECRIS plasmas are density limited in a pure electromagnetic plasma ignition scheme. To overcome this limitation, Bernstein Waves [14] heating has been recently attempted in nuclear fusion devices like Stellarators and Tokamaks. BWs, which are waves of matter, propagate in plasmas of whatever density and can be only produced via inner-plasma electromagnetic-to-electrostatic waves conversion through conversion processes called XB or OXB mechanisms [15,16].

In a pilot experiment recently performed at INFN-LNS, with a small-size prototype, densities from ten to twenty times above the cutoff have been obtained, once reached a power threshold; correspondingly, the plasma emits intense fluxes of X-rays up to 30 keV of energy, manifesting the formation of a 3D, "typhoon-shaped" plasma vortex [17–20].

At INFN-LNS several innovative approaches and numerical tools are now available in order to further improve ECRIS performances, especially based on advanced plasma modelling in order to design innovative RF launchers. The novel idea, as devised in PANDORA, is that ECRIS plasma traps are suitable for non-conventional applications due to ECRIS compactness, to the high energy content of the plasma, to the MHD stabilization under proper magnetic field configurations.

## 3 The Physics case:

### 3.1 Nuclear Astrophysics

The investigation of nuclear phenomena like fusion reactions, or radioactive decays for astrophysical purposes, is growing in interest in the last decades: issues yet unsolved on stellar and primordial nucleosynthesis or neutrino production, in fact, require precise reaction rates or radionuclides lifetime measurements [21]. The development of innovative experimental techniques, such as nuclear reaction measurements on ionized samples, will be beneficial for the study of nucleosynthesis in stars, as it will open the way to challenging measurements in the field (e.g. Electron Screening effects in low-energy nuclear reactions). In the recent past, several strategies have been attempted in order to modify nuclear decay rates by changing environmental conditions [22, 23]: this research field has many astrophysical implications and technological applications (energy production, nuclear waste transmutation, etc.). In this context, it has been demonstrated by several experiments performed at GSI on highly-charged ions that Electron Capture (EC) decay probability can be dramatically modified by the atomic electron configuration [23–26]. As an example, in case of H-like and He-like $^{140}$Pr [23] although the number of atomic orbital electrons was reduced from two in $^{140}$Pr$^{57+}$ ions to only one in $^{140}$Pr$^{58+}$ ions, the EC decay rate increased by a factor of 1.5. These evidences have opened up a new frontier in the investigation of EC decays in multi-charged ions and it has extraordinary implications in nuclear astrophysics and neutrino physics. In fact, terrestrial measurements involve, up to now, neutral species, while both nuclear reactions and decays involved in nuclear astrophysics should be investigated in peculiar plasma environments. Indeed, plasma is a quite common state of matter in our universe, having properties unlike those of the other states. As an example, the neutral $^7$Be atoms are radioactive and decay with a half-life of about 53 days to $^7$Li atoms via the only energetically allowed decay channel, the orbital EC decay. Conversely, ionized (in plasma) $^7$Be EC decay probability is unknown, representing a crucial parameter both in stellar and primordial nucleosynthesis models as reported below.

### 3.2 Solar nucleosynthesis and neutrino flux

Ionized and fully stripped $^7$Be are key participants in the pp chain, responsible of the main energy production mechanism in our Sun. In this scenario models predict that the main decay channel of $^7$Be in the Sun's interior is the free electron capture, but orbital electrons can increase the decay probability[27]. In this framework theoretical calculations show that about 20% of $^7$Be in the core of the Sun might have a bound electron [27]. As a consequence, a shorter/longer lifetime of $^7$Be in the Sun would modify the $^7$Be neutrino lines emission and decrease/increase the probability of its destruction via the $^7$Be(p,$\gamma$)$^8$B capture reaction, in turn changing the yield of the solar neutrino flux in the high energy range. Shown in figure 2 the actual neutrino-flux model predictions by [28, 29].

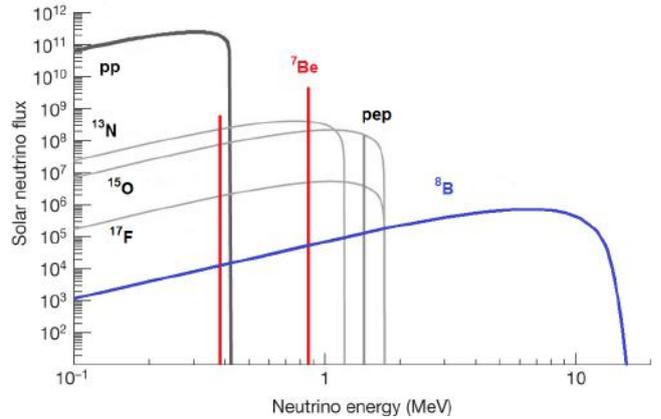

**Fig. 2.** Estimated neutrino fluxes coming from the Sun [29].

### 3.3 The Cosmological Lithium Problem

One of the most puzzling problem in Big-Bang Nucleosynthesis (BBN), a cornerstone of Big-Bang theory, concerns the so called "Cosmological Lithium Problem" (CLP) [30]. It is striking that BBN model reproduces the observations of primordial abundances for all isotopes except $^7$Li, for which it overestimates the relative experimental abundance by about a factor of 3 [30–32].

An explanation involves large systematic uncertainties in the Nuclear Physics inputs of the BBN calculations, in particular in the cross-sections of reactions leading to the destruction of $^7$Be [31], whose decay is responsible for the primordial abundance of $^7$Li, anyhow in a recent work by the n_TOF collaboration at CERN [32] the possibility of explaining the CLP in terms of an enhancement of the destruction channel $^7$Be(n,$\alpha$) has been ruled out. Again, $^7$Be decay in primordial nucleosynthesis happens in a quite peculiar plasma environment and any modification of the decay probability could significantly affect the BBN network, deeply modifying the whole scenario.

### 3.4 $^7$Be in-plasma decay-rate measurement

The study of $^7$Be EC decay probability in a real plasma environment will be competing and complementing with respect to the in-ring experiment now in start-up phase at ISOLDE-CERN (Switzerland) [33] by using the TSR ISOLDE setup. In our case the decay probability will be measured as a function of charge-state distribution in a long-term stationary dynamic-equilibrium. To this aim it is mandatory to have, high plasma density and a stable and reproducible plasma conditions, an accurate on-line monitoring of in-plasma $^7$Be charge state distribution (CSD), an easy and reliable tuning of plasma electron temperature in order to explore a wide range of charge state distributions and finally a low $^7$Be consumption rate due to the small quantities available. Such requirements will be fulfilled by the innovative design of the new high-performance PANDORA ECRIS plasma trap in combina-tion with the new plasma diagnostics and Charge Breeding injection which will be discussed in the following sections.

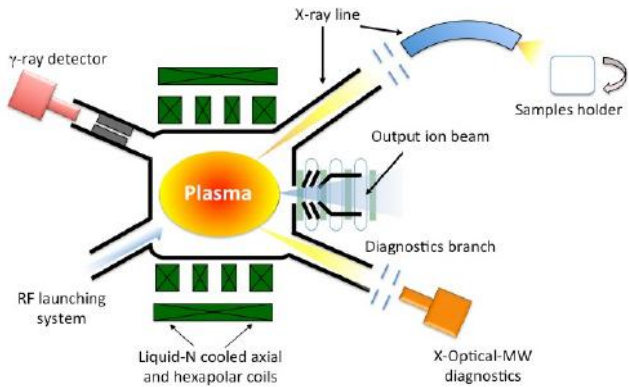

**Fig. 3.** General sketched layout of the proposed machine and accessory lines for Nuclear Decay observation and X-radiation extraction.

By our approach EC decay-rate to the $^7$Li first excited state can be measured by using a large efficiency HPGe detector placed in a collinear geometry with respect to the ECR plasma trap. Decay will be tagged by detecting 477.6 keV $\gamma$-rays. In figure 3 is sketched the general setup, based on the above-mentioned B-minimum plasma-trap structure, including plasma diagnostics and suitable accesses for $\gamma$-ray spectroscopy. Concerning the plasma, the magnetic field, the RF power and the pumping frequency will be tuned to ensure plasma parameters stable for several hours. The RF power will be kept as high as possible to ensure the maximum efficiency in fast Be ionization and to sustain the maximum plasma density. The magnetic trap will be kept as short as possible to reduce X-ray generation above 400 keV (i.e. below the characteristic line of the $^7$Be emitted $\gamma$-ray) and to select electron energies able to provide different charge states distribution of $^7$Be. The mechanical set-up will be carefully optimized according to the requirement of shielding the detection system from any other source of $\gamma$ radiation (e.g. atomic $^7$Be deposited on the ECR walls): a specific lead collimator is being designed. $^7$Be isotope is available in micro quantities ($\sim$ 1-10 $\mu$g)[34], consequently the in-trap injection and consumption rate must be optimized. A 3D modeling of the plasma (3D density, temperature and charge state maps) will be crucial to simulate 477.6 keV $\gamma$-rays emission, in order to carefully determine the best detector placement.

### 3.5 s-process branch-points in PANDORA

Among the various processes responsible for the formation of the heavy elements in stars, the slow neutron capture process (s process) involves mostly stable isotopes. Therefore, the relevant nuclear physics data can be determined by experiments involving n-capture reaction and by measuring the decay probability at the so called s-process branch-point [21]. With this data set, s-process nucleosynthesis offers an important test-ground of models in the late stages of stellar evolution, which are supposed to be the s-process site. At our knowledge no in-plasma measurement has been performed in order to determine the decay probability for nuclide's of the s-process branch-point. One interesting case, for instance, concerns $^{151}$Sm [35](see figure 4). Terrestrial $^{151}$Sm has a half-life of 90 years while in astrophysical plasma it is supposed to be reduced to just few years.

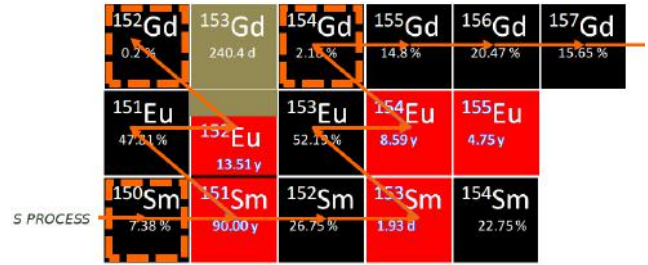

**Fig. 4.** Scheme of nuclear decay chains in s processes playing a crucial role in stellar nucleosynthesis.

Consequently the competition between slow neutron-capture and decay at the branch-point, gives information about the average neutron density, temperature and matter density in the stellar plasma during the s-process nucleosynthesis.

### 3.6 Cosmochronometers in PANDORA

Once determined the nucleosynthesis rate and by using the abundances of long-lived radioisotopes, one can estimate a lower bound for the age of the galaxies. This clarify the fascinating concept of cosmochronometers like $^{232}$Th/$^{238}$U and $^{187}$Re/$^{187}$Os pairs, introduced by Clayton [21]. However, one uncertainty in the calibration of these cosmic clock has been pointed out by [36]: i.e. the nuclides may become highly ionized in the hot plasma of a star, and bound-state beta-decay in $^{187}$Re/$^{187}$Os pair may decrease the half-life from 42.3 Gyr by more than 9 orders of magnitude. According to this calculations PANDORA offers a suitable hot-plasma environment where to measure, for the first time, this huge half-life variation. Obviously a plethora of new exotic ionic species decaying by bound-state beta-decay could be discovered in a long term experimental campaign, establishing a solid basis for a new conceptual and methodological trend in Nuclear and Plasma Physics, towards a cross fertilization among atomic, nuclear and astro physics.

### 3.7 Cosmic Magnetic Fields

Magnetic fields are expected to play a fundamental role in stellar physics across the entire Hertzsprung- Russell (HR)

diagram. They have been invoked to explain phenomena such as non-thermal radio emission from O-B stars and jets in young stellar objects; they are responsible for activity in the Sun and in late-type stars, and can reach huge strengths in degenerate stars. However, the importance of magnetic fields is not limited to the understanding of particular stellar phenomena. They are fundamental in the modeling of stellar structure and evolution, because they are able of modifying the observational variables changing the location of stars in the HR diagram and strongly interacting with mass motions. The largest effect of magnetic fields on stellar photospheres is observed in more than 10% of the A-type stars on the Main Sequence (MS): the so called Ap stars. Inhibiting any photospheric motions, magnetic fields allow radiative diffusion to strongly modify the chemical composition profiles [37]. Over- and under-abundances, by up to several dex with respect to the Sun, together with Zeeman splitting [38] affect the line opacity and the whole atmospheric structure and modify the overall flux distribution and Balmer line profiles. Effective temperatures may be in error by up to several thousands of degrees and gravities by some tenths of dex [39], when classical relations are used. As a consequence magnetic stars are often incorrectly located in the HR diagram. For massive stars, Maeder and Meynet [40] have shown that the inclusion of rotation and magnetic fields changes the core mass fraction, reducing the MS lifetime. Magnetic fields affect the depth of the convective layers and the over-shooting region in outer layers, D'Antona et al. [41] have shown that evolutionary tracks of T Tauri stars are 1000K cooler when a mere 45G surface field is included. Moreover, magnetic fields drive stellar winds, modifying mass loss [42] and angular momentum of a star [43]. Both phenomena strongly affect stellar evolution.

A direct measurements of cosmic magnetic fields can be based only on the Zeeman effect as observed at high resolution spectropolarimetry. However, with the exception of few, mainly solar, spectral lines, effective Landé factors are estimated in the L-S approximation and these factors can be 50% different with respect the experimental values [44]. Then, it looks that the knowledge of one of the most important parameters in astrophysics is limited by the lack of experimental values of Landé factors.

The magnetized plasmas we plan to use in PANDORA are a fundamental experimental environment to perform a wide range of measurements of astrophysical interest. Particularly, we plane to obtain experimental values for the second order Landé factors, these are the equivalent of the effective Landé factors for the $\pi$ components, and convert the observed linear polarization across spectral lines in the intensity and orientation of the transverse component of a magnetic field.

Calibrations or benchmark measurements of atomic parameters will not be limited to the Landé factors. Stark broadening and isotopic shifts of spectral lines will be also routinely determined. Further studies can be conducted on the issue of plasma radio emission [45] with the aim to characterize what we observe from highly magnetized stars [46,47].

### 3.8 High Energy Astrophysics

Started few decades ago, High Energy Astrophysics is rapidly growing from the observational and theoretical points of view. In the forthcoming few years, High Energy Astrophysics will be driven by the crossmatch of data collected with the most specialized ground and space-based telescopes covering the: Radio and sub-millimiter (JVLA, SKA, ALMA), Optical and Infrared (E-ELT, JWST), X-rays (Chandra, SWIFT, NuStar) and $\gamma$-rays (INTEGRAL, Fermi). These "classical" information will be complemented with cosmic rays (Auger, CTA), neutrinos (IceCube, Km$^3$NeT) and with gravitational wave astronomy (LIGO, VIRGO). In such a context, PANDORA will be one of the few facilities able to support the High Energy Astrophysics by the definition of a complete database of spectral lines in the X-ray domain. The most recent related example is the 3.5 keV X-Ray spectral line observed towards galaxies. It has been suggested [48,49] that this X-ray line can be produced by the decay of the dark matter particle candidates: the sterile neutrinos. Differently, Gu et al. [50] ascribed the 3.5 keV X-ray to a charge exchange emission due to $S^{16+}$ ion. A scenario experimentally confirmed by Shah et al. [51] within an electron beam ion trap.

## 4 The experimental scenario: the CSD tuning

The densities of the multiply-charged ions to be confined for providing meaningful information concerning decay rates in plasmas will require a Magneto Hydro Dynamically stable regime of trapping. Figure 5 shows preliminary theoretical estimations about plasma CSD obtained via the model based on balance equations:

$$\frac{\partial n_i}{\partial t} = \sum_{j=j_{min}}^{i-1} n_e n_j \left\langle \sigma_{j \to i}^{EI} \nu_e \right\rangle + n_0 n_{i+1} \left\langle \sigma_{i+1 \to i}^{CE} \nu_{i+1} \right\rangle - n_0 n_i \left\langle \sigma_{i \to i-1}^{CE} \nu_i \right\rangle - \sum_{j=i+1}^{j_{max}} n_e n_j \left\langle \sigma_{j \to i}^{EI} \nu_e \right\rangle - \frac{n_i}{\tau_i} = 0$$

where $n_e n_j \left\langle \sigma_{j \to i}^{EI} \nu_e \right\rangle$ denotes the number of ions per second produced by electron impact from the species at charge state $j > i$; $n_0 n_{i+1} \left\langle \sigma_{i+1 \to i}^{CE} \nu_{i+1} \right\rangle$ is the number of ions generated by charge-exchange collisions with neutrals; $n_0 n_i \left\langle \sigma_{i \to i-1}^{CE} \nu_i \right\rangle$ denotes the number of ions at charge state $i-1$ produced by charge-exchange recombination; $\sum_{j=i+1}^{j_{max}} n_e n_j \left\langle \sigma_{j \to i}^{EI} \nu_e \right\rangle$ gives the number of ions at charge $i$ ionized by the electrons and shifted to charge state $j > i$; $\frac{n_i}{\tau_i}$ represents the plasma losses due to diffusion to the walls or extracted current from plasma. Free parameters are typically $n_e$, $\tau_i$ and $f(v_e)$. The figure 5 shows the trends of the different charge states abundances for $^7$Be as a function of temperature, assuming the plasma confinement time as a parameter. The simulations support the underlying idea of PANDORA, i.e. that plasma produced by the proposed machine can simulate stellar-like conditions in terms of Beryllium charge states. It can be seen that the variation of the plasma temperature (tuning of RF power) or the adjustments in the plasma confinement time (obtainable by acting on the magnetic field), modify in a significant way the relative amounts of the different ions, thus permitting to correlate decay times with the electronic configuration of the ions. These preliminary data motivate why does the flexibility of the magnetic field is a mandatory requirement when designing the new plasma machine prototype: it has to be adjusted in order to promote different plasma CSD configurations.

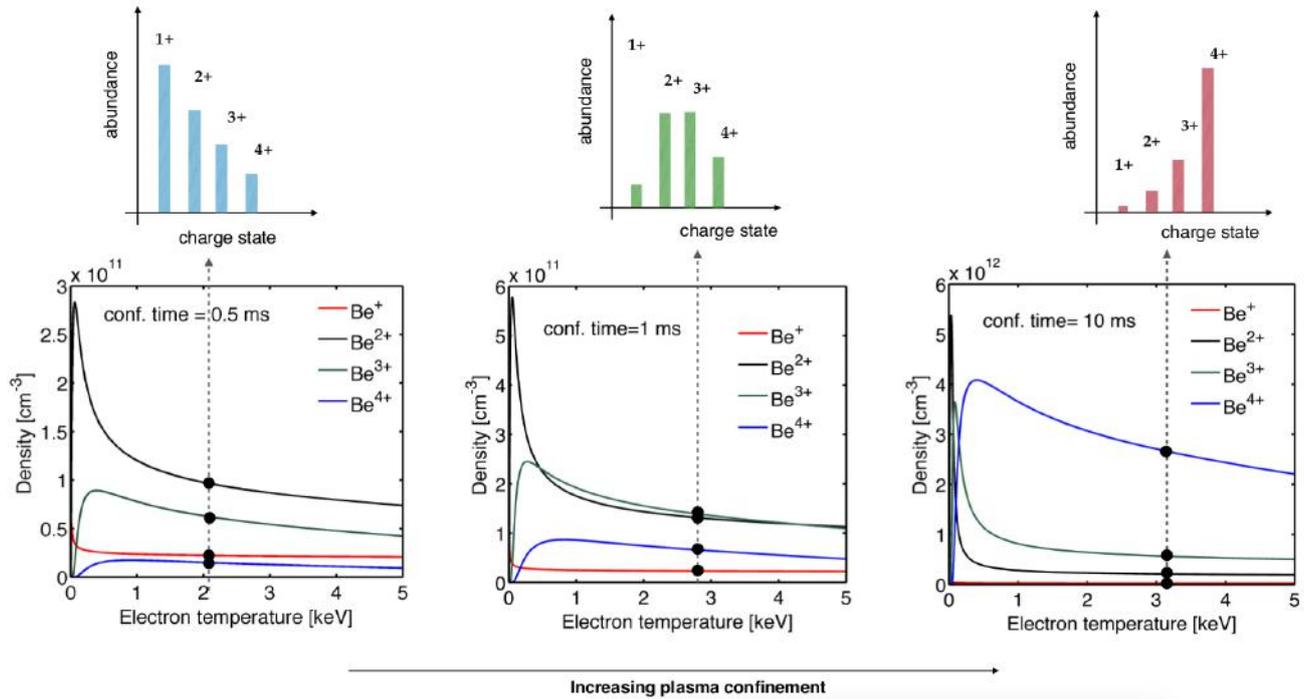

**Fig. 5.** Theoretical calculations concerning the abundances of $^7$Be charge states vs. in-plasma density, temperature and ion average lifetime.

## 5 The PANDORA plasma-trap

As already stressed, the densities of the multiply-charged ions to be confined for providing information on any long-term physical phenomenon occurring in plasmas will require a MHD stable regime of trapping (i.e. hydro-dynamical turbulences suppressed in a low-$\beta$ values of the plasma-magnetic system). Hence, MHD stability will be a crucial experimental condition, since a stationary plasma state is needed in order to correlate the investigated in-plasma phenomena with plasma observables, namely: average charge state, temperature and density.

In Table 1 the main characteristics of the PANDORA plasma machine and the expected performance are summarized. We plan to exploit (for the first time in a linear compact device) the inner plasma mode-conversion (XB or OXB mechanisms [15,16]) exciting electrostatic Bernstein waves (EBW) up to 18 GHz, thus producing multi-charged ions with an operative density approaching $10^{14}$ cm$^{-3}$, like in the inner layers of a stellar atmosphere. When used as ion source, it will be one of the most performing worldwide. Simplified magnetic structures (Simple Mirror or Off-Resonance) will be tested for EBW generation, and the results compared in the two different conditions: ECR-heating and EBW-heating. When heated by EBW, the plasma will score the best performances in terms of UV/soft-X ray emission (see Table 1). Up to early 2000, ECRIS plasmas were considered as "sponges" absorbing in some way almost all the RF power injected into the resonator (i.e. the plasma chamber). However, the single-pass RF energy absorption efficiency is rather poor, and it is still difficult to drive energy deposition to specific parts of electron population in the phase space. In PANDORA we propose a "microwave absorption optimization-oriented" design based on a highly controlled electromagnetic radiation by the RF launcher system. The launcher will be very flexible, allowing frequency tuning, polarization control, power deposition localization (high directivity). The most important requirement is the beam direction variation as in a quasi-optical scheme (see figure 6); different strategies will be investigated: this is mandatory in order to establish the electron Bernstein wave formation and damping, due to the mode conversion from electromagnetic to an electrostatic wave: it strongly depends on the polarization and direction of the incident wave.

| Technological and Physical Characteristics | EXPECTED VALUES | CHALLANGES |
|---|---|---|
| **Operative frequencies** | 12-18, or 12 + 18 GHz | Use of lower frequency for EBW, forming an overdense plasma |
| Maximum RF power | 2+1 kW @ 12 + 18GHz | |
| B field values (max-inj/min/max-rad) | 2/0.3/1.6-1.6 T | |
| Type of magnetic structure | L-Nitrogen cooled copper coils (4 axial and 6 for hexapole radial field) | Reach high currents in the coils by cooling copper with liquid nitrogen |
| Operative pressure | Range 10$^{-7}$ mbar | |
| Plasma volume | 1-2 dm$^3$ (tunable) | |
| Total extracted current | > 10 mA at 40 kV in B-min mode, >100 mA in off-resonance mode @ 75 kV | off-resonance at high freq. never attempted in the past |
| Max plasma chamber diameter/length | 200/500 mm | |
| Elements to be ignited | Gas & Metals | |
| **Operative plasma density** | 10$^{11}$-10$^{13}$ cm$^{-3}$ | High plasma density comparable to Superconducting ECRIS at 28 GHz |
| Typical hot plasma temperature | 1-150 keV | |
| Expected X-rays energy endpoint | 200-800 keV | |
| **Expected X-ray flux** | 10$^{18}$-10$^{19}$ counts/sec/4π (0.1<E<1keV) 10$^{14}$-10$^{15}$ counts/sec/4π (1<E<5 keV) | High fluxes of radiation never extracted in the past in a table top machine |

**Table 1.** Technical characteristics of the proposed new plasma source.

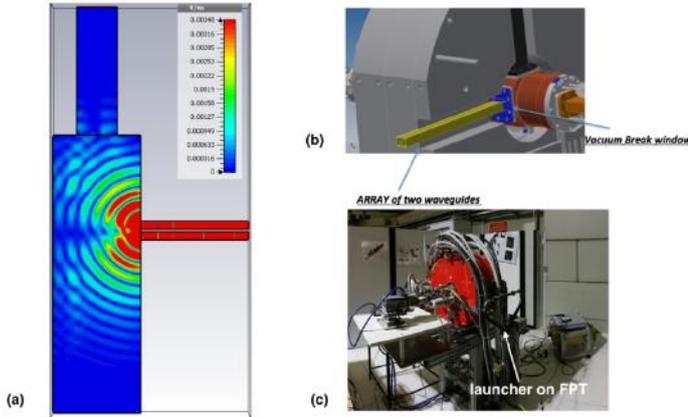

**Fig. 6.** Quasi-optical RF launching setup installed in the Flexible Plasma Trap (FPT): a) simulation of the Electromagnetic Wave in-trap injection; b) mechanical drawing of the wave guides inside the PST; c) photo of the FPT showing the launcher [52]

According to the previous experimental constraints concerning multi-charged ion half-life and spectropolarimetric measurements, two demanding tasks must be accomplished by the setup:

i) PANDORA must be able to perform space-resolved measurements of the in-plasma Charge State Distribution of ions.

ii) It is mandatory to optimize the injection efficiency of the suitable radioactive elements inside the magnetized plasma. In fact, in most of the cases, such species are available in micro quantities. These subjects will be treated in the following subsections.

### 5.1 Plasma diagnostic

The radioactive activity of the injected species must be correlated to the in-plasma charge state distribution. Previous plasma investigations [53] have shown that the plasma energy content is not uniformly distributed inside the trap, therefore space-resolved diagnostics methods are required for a complete characterisation. In table 2, the experimental techniques needed for a complete plasma characterization are listed.

The Laboratori Nazionali del Sud-INFN(LNS-INFN) plasma-physics group, supported by the INFN 5th National Commission, has studied, developed and implemented in the last decade several advanced plasma-diagnostics techniques[53–57]. Hence, of the ones listed in Table II, some are already available at LNS and they work routinely for plasma characterization, namely the ones labelled as A1-A3, B2, B3, E1. C1 and C2 are under test-phase at LNS. Among the others, it is worth mentioning the A3 technique (the X-ray pin-hole camera) that has allowed to characterize the plasma morphology and to perform space resolved spectroscopy (thus evidencing the local displacement of electrons at different energies, as well as of plasma ions highlighted by fluorescence lines emission) versus the main tuning parameters such as the pumping wave frequency and the strength of the confining magnetic field.

| | |
|---|---|
| **A. Warm & Hot electrons Temperature** | • A1 – Continuous and characteristic X radiation E<30 keV measured by SDD detectors;<br>• A2 – Hard X-rays (E>50 keV, up to hundreds keV) by large volume HpGe detectors;<br>• A3 – X-rays (1E>20 keV) pin-hole camera with high energy resolution (around 150 eV) for space resolved X-ray spectroscopy; |
| **B. Cold Electron Temp. & Density** | • B1 – Space Resolved Optical Emission Spectroscopy (space resolution less than 100 μm and spectral resolution of about $10^{-3}$ nm in the range 200-900 nm;)<br>• B2 – Line integrated density measurement through microwave interferometry;<br>• B3 – Faraday-rotation diagnostics (horn antennas coupled to polarization meter); |
| **C. Ion Temperature** | • C1 – Measurement of X-ray fluorescence lines broadening through high resolution ($\Delta\lambda/\lambda = 10^{-3}$) X-ray spectroscopy, by using doubly curved crystals coupled to polycapillars;<br>• C2 – Space resolved measurements are possibile with a Polycapillar+doubly-curved-crystal+CCD (X-ray sensitive) camera in a "pin-hole method" scenario; |
| **D. On-line Charge State Distribution (CSD)** | • D1 – Space Resolved Optical Emission Spectroscopy (space resolution less than 100 μm and spectral resolution of about $10^{-3}$ nm in the range 200-900 nm;)<br>• D2 – X-ray fluorescence lines *shift* through high resolution ($\Delta\lambda/\lambda = 10^{-3}$) X-ray spectroscopy (curved crystals + polycapillar);<br>• D3 – Space resolved measurements: Polycapillar+doubly-curved-crystal+CCD (X-ray sensitive) camera in a "pin-hole method" scenario; |
| **E. Off-line CSD measurement** | • E1 – Obtained after ion beam extraction, through a magnetic dipole (magnetic mass spectrometer) already installed at the INFN-LNS beam transport line; |

**Table 2.** Diagnostics techniques to be used for plasma characterization and plasma monitoring during nuclear decays observation.

Figures 7 and 8 illustrate the summary of the recently performed measurements showing the different displacement of electrons in the various energy domains inside the plasma core, and along the branches of the six-arms (only two are visible in the figure) shaped B-minimum magnetic field configuration. Another example of non-intrusive diagnostics is the sweep-frequency microwave interferometer VESPRI, developed since 2013 at LNS. The interferometer (shown in figure 9) has performed the first non-intrusive and line-integrated measurement of plasma density [55].

The next step will consist in performing a 1D profilometry of the density, along the line-of-sight. In combination with transversal 2D imaging and space resolved X-ray spectroscopy, it will be an important tool for reconstructing the 3D structure of the plasma.

The completion of the PANDORA "diagnostics toolbox", that will able to perform the on-line characterization of the CSD, will be achieved by a valuable opportunity in the framework of the strict synergy activated with the INAF Istituto Nazionale di Astrofisica, Osservatorio Astrofisico di Catania (INAF-OACT).

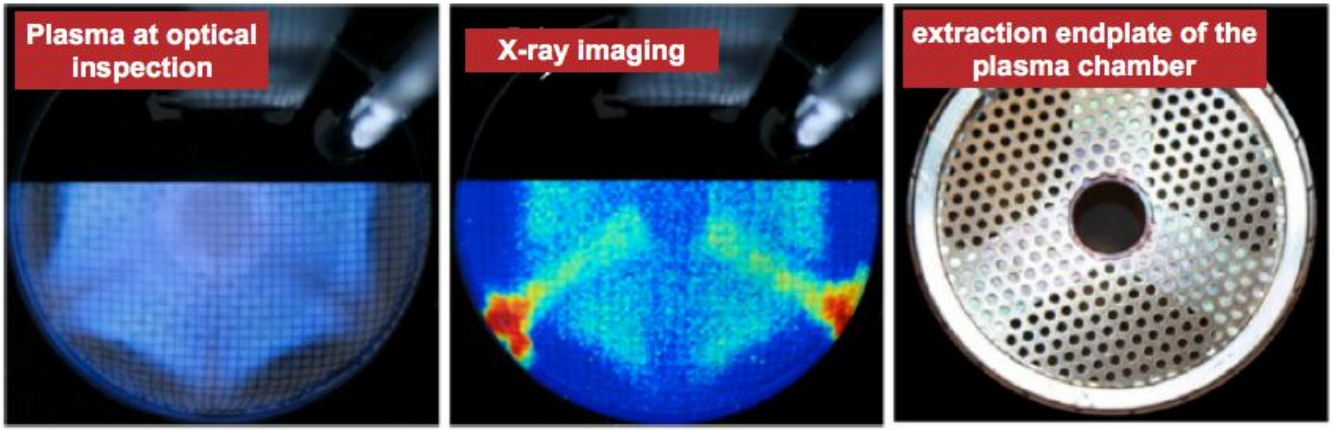

**Fig. 7.** Results collected during X-ray space resolved spectroscopy experiments on a magnetically confined plasma in B-minimum configuration: plasma morphology in the visible and X-ray domains.

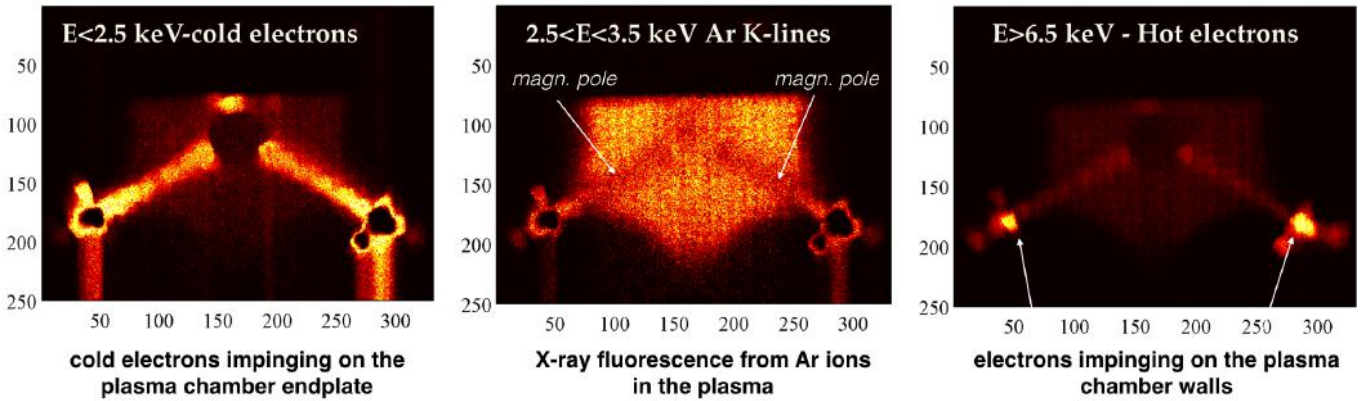

**Fig. 8.** Results collected during X-ray space resolved spectroscopy experiments on a magnetically confined plasma in B-minimum configuration: different slices in the X-ray domains [55,56]

In fact the spectrograph named SARG (Spettrografo Alta Risoluzione Galileo) [59], one of the worldwide most powerful spectrograph for astronomy, will be installed at LNS, where it will be coupled for commissioning to the F.P.T.(Flexible Plasma Trap, see figures 6 and 10) facility [60].

SARG allows to reach very high resolution: R=160.000 in the range: 370-900 nm, i.e. fractions of Angstroms, which are suitable to discriminate plasma-emitted spectra coming from different CSD. Some simulations have already been performed and results are shown in figures 11 and 12. In particular figure 11 depicts the expected spectra for an oxygen plasma, having two different bulk electron temperatures and populated by several charge states.

The variation of the temperature from 10 to 15 eV causes the upshift of the CSD towards higher charge states. In any case, for both the temperatures the resolution of the SARG is high enough to discriminate the different charge states of the oxygen, labeled as O-II O-V in the plot. Taking advantage of the isotopic shift of spectral lines, see figure 12 [61], the high resolution will allow to discriminate in-plasma isotopes with high precision, thus offering an exciting additional opportunity to evaluate the transmutation of radionuclides directly inside the plasma trap.

### 5.2 In-trap injection: the Charge-Breeding technique

Before describing and justifying the Charge Breeding technique to be used in PANDORA, we will briefly comment on the techniques currently used to feed neutral atoms in ECR ion sources for stable beams production, as comparison. In the case of solid elements, two techniques are traditionally used, i.e. the resistive oven and the sputtering, both presenting several drawbacks for PANDORA. Resistive ovens consist in a crucible with a very thin filament wrapped on it: they are usually very small, and are mounted inside the plasma chamber, beyond the maximum of the magnetic field at injection. The sputtering technique is usually used for refractory elements: in this case, a target of the element to be sputtered is mounted on a rod inside the plasma chamber, at more or less the same position as an oven. The rod is polarized negatively at some kVs with respect to the plasma chamber (sputtering voltage), so as to accelerate the positive ions leaking from the plasma, producing sputtering.

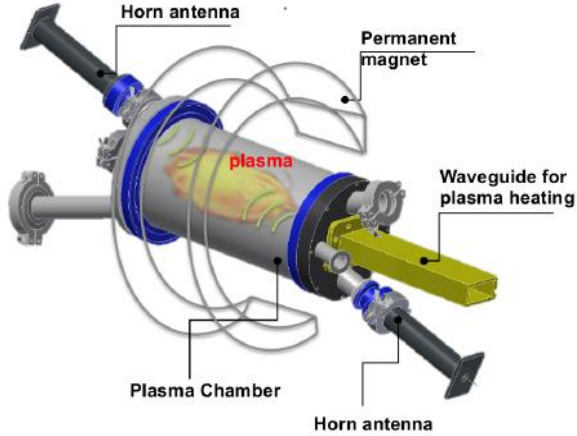

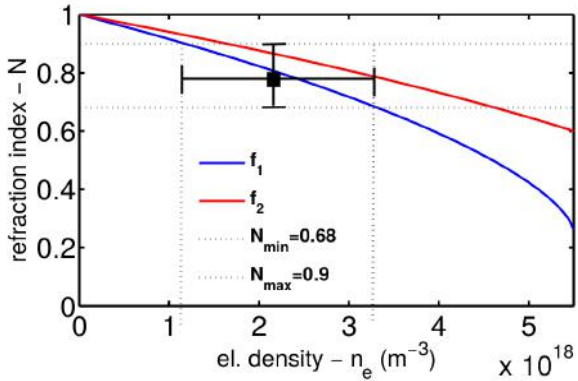

**Fig. 9.** First measurement of plasma density in a compact plasma trap (bottom) and the relative experimental setup implemented at INFN-LNS for the installation of the VESPRI interferometer (top) [57,58].

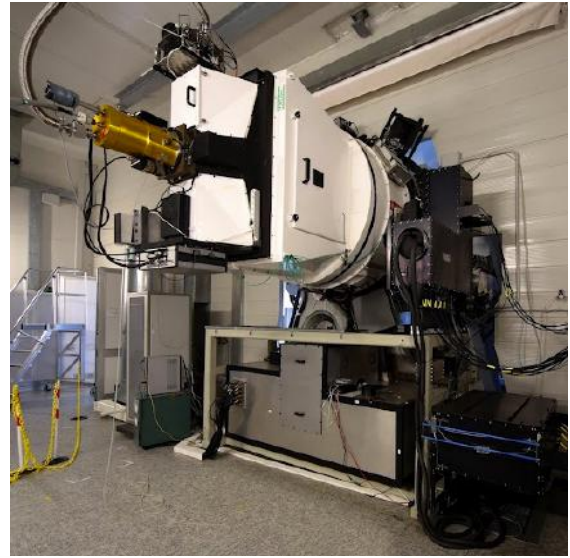

**Fig. 10.** Photo showing the SARG spectrograph (bottom part) coupled to DOLORES (Device Optimized for the LOw RESolution) a focal reducer instrument installed at the Nasmyth B focus of the TNG.

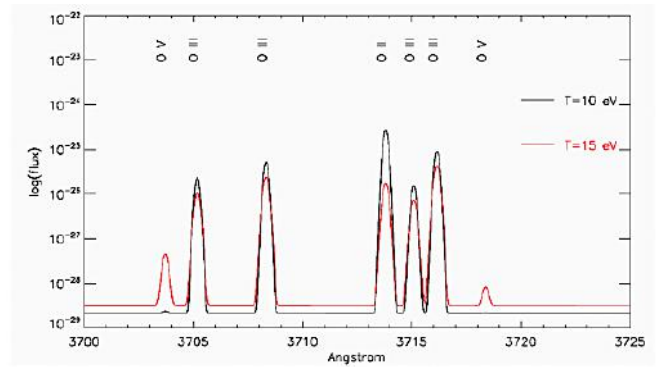

**Fig. 11.** Simulated optical emission spectrum from an oxygen plasma as measured by SARG.

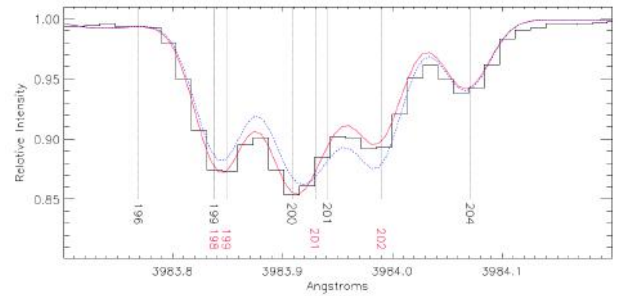

**Fig. 12.** Hg II 3984 Å spectral line (black) of HD216494 is matched with a mercury isotopic mixture (red), slightly different than the terrestrial one (blue).

In both cases the sample preparation is laborious, and imply hazardous handling of the radioactive specie for a long time. In the case of the oven, the estimated temperature to create vapors of $^7$Be is around 1200 °C (at this temperature its vapor pressure is 1 Pa and Beryllium sublimates): this implies the use of particular materials to build the oven, so as to ensure long term operations at high temperature. More, it is widely known that metallic vapors are captured with low efficiency (10%-20%): this would imply that a significant amount of $^7$Be atoms would condensate on the plasma chamber wall, leading to a high and diffuse contamination. This would produce a gamma background not distinguishable from the signal coming from the core of the plasma; moreover, if the heating fil-ament breaks during operation, the evaporation does not stop immediately, due to the thermal inertia of the oven itself: i.e. it is not possible to stop suddenly the evaporation. The sputtering technique does not present this last inconvenience, but the neutral particles created have energies of several eVs, leading again to a low capture efficiency by the plasma. Furthermore, the sputtering yield is directly influenced by the plasma parameters (the projectiles are plasma ions), so it is not possible to tune independently the sputtering process and the plasma characteristics. Finally, the high voltage applied to produce sputtering can perturb the plasma, leading to plasma instabilities and ion losses. These last issues are real concerns when radioactive elements are involved.

To avoid the inconveniences described above, and optimize the injection and capture of the radioactive ions inside the plasma-trap, we plan to apply the Charge Breeding Technique (CBT): a simplified scheme is shown in figure 13. By this technique, first a radioactive 1+ beam is produced in a hot ion source, like the ones used in ISOL facilities, and then extracted: for the sake of clearness, figure 14 shows the source will be employed for the SPES project [62,63]. Then, the extracted beam is transported through an electrostatic low energy beam-line, and it is injected into an RFQ cooler, in order to lower its emittance (down of a factor of 10 depending on the element) and its energy spread ($\sim 1$ eV), very important parameters for an optimum injection in the following trap. Inside the RFQ cooler, ions are decelerated to 100-200 eV and propagate in a linear channel, with a strong radial focusing due to a radio-frequency quadrupole that induces transverse confinement through the electric field:

$$E_{rf} = -\nabla \phi_{rf} \quad \phi_{rf} \cong \frac{V_{rf}(z)}{r_0^2}\left(x^2 - y^2\right)\cos\omega t$$

with z the propagation axis, $\omega$ the angular frequency (in the order of MHz), $V_{rf}$ the peak $RF$ voltage at electrodes, and $2r_0$ the distance between facing electrodes. Under the effect of the $RF$ transverse field, particles undergo radial oscillations at the frequency:

$$\omega_m \equiv \frac{q\omega}{\sqrt{8}}, \quad q = \frac{4eV_{rf}}{m\omega^2 r_0^2}$$

where $q$ is a dimensionless parameter known as Mathieu parameter, whose value has to be $0 < q < 0.91$ for a stable motion.

By keeping a pressure of around 1 Pa of a light gas (usually He) inside the cooler, collisions with neutral particles produce a friction force that damp transverse oscillation and lower the longitudinal energy spread. To drag cooled ions towards the extraction, a longitudinal electric field $E_z (\sim 100 V/m)$ is usually superimposed to the quadrupolar field. Figure 15 shows the RFQ cooler developed for the SPES project [64], composed by three main section: injection, RFQ-cooling and extraction. The same figure shows a simulation carried out by the tracking code SIMION, with a $^{133}Cs^{1+}$ beam injected, cooled (transverse oscillations decrease inside the cooler) and then extracted as a high quality 1+ beam. The same code will be used to study the parameters in order to optimize $^7Be^{1+}$ ion cooling. The ion beam extracted from the cooler can then be injected into the B-minimum trap (playing the role of a CB): as soon as the 1+ beam approaches the CB, it is decelerated by its electrostatic potential (the charge breeder is at high voltage), and feels a reflecting force due to the magnetic fringing field. By properly regulating its injection energy (in the order of few eV), the beam is able to overcome the maximum of the magnetic field and the plasma potential, and penetrates in the plasma core. Here, injected ions undergo collective Coulomb collisions, leading to thermalization and consequent stepwise ionizations. The charge breeding technique overcomes the drawbacks of the two injection techniques previously described, in fact: i) the injection process can be stopped at any moment by acting on the 1+ source, or its downstream Faraday Cup, or any other element of the beam line; ii) the capture efficiency is usually much higher than the other techniques (it can reach 90% for gaseous element). iii) Finally, the contamination by lost particles is reduced to the well-known six strips along the plasma chamber wall, created by the magnetic field loss lines (see for example figure 16): such strips are very well localized and of small extension, so the gamma detector could be properly collimated to not record gamma produced by the lost $^7Be^{1+}$ ions.

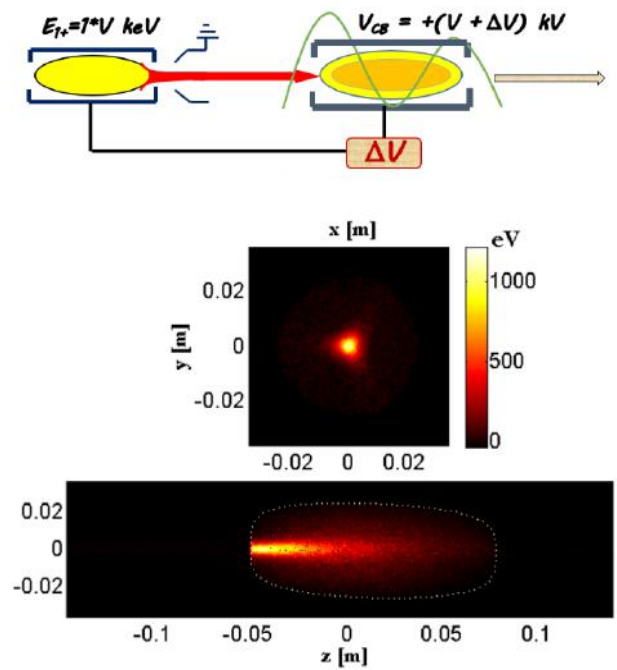

**Fig. 13.** Scheme of the Charge Breeding Technique: a radioactive beam is extracted from a hot $1^+$ source, and injected into the charge breeder to enhance its charge state (top figure). Spatial distribution of the energy released by the injected particles inside the plasma on the xy plane and xz plane (central and bottom figures).

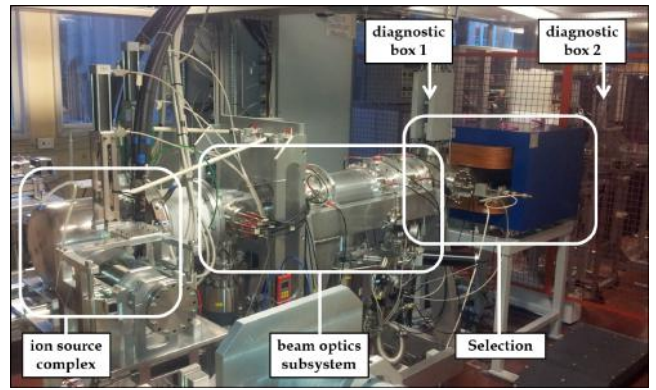

**Fig. 14.** Layout of the hot $1^+$ source and its beam-line employed in the SPES project.

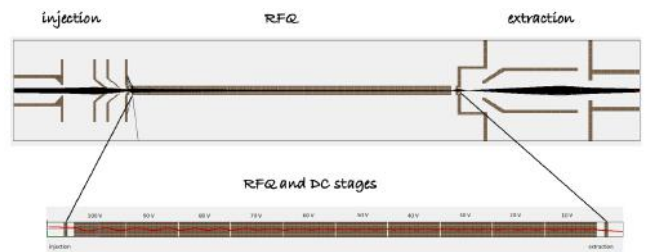

**Fig. 15.** Scheme of an RFQ cooler (top figure); simulation of the cooling of a $^{133}Cs^{1+}$ carried out with the tracking code SIMION (bottom figure).

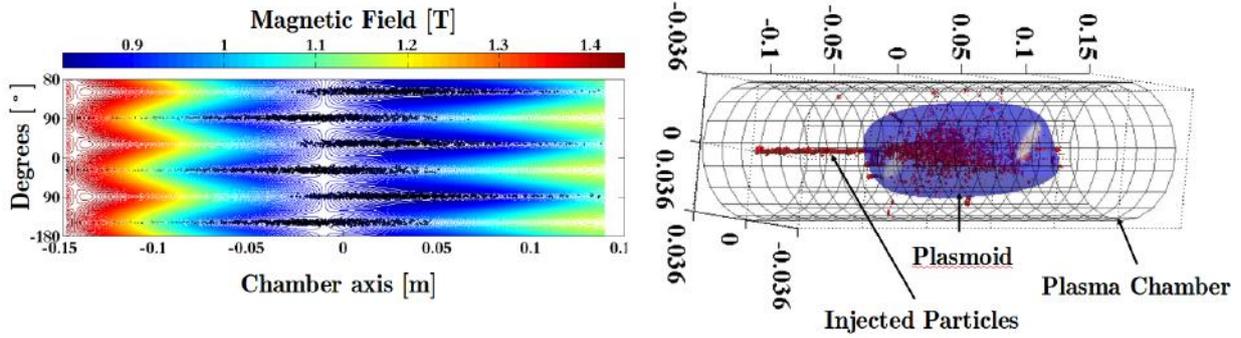

**Fig. 16.** Left side: distribution of ion losses on the plasma chamber wall. Right side: distribution of the particles captured by the plasma[65].

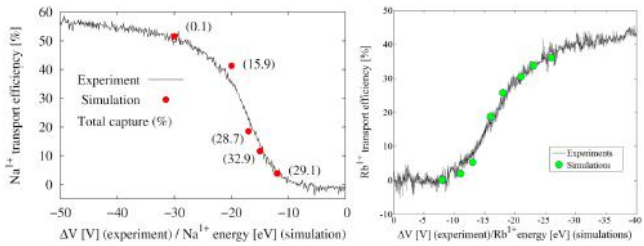

**Fig. 17.** Comparison between numerical and experimental $1^+$ efficiencies for Na (left)[66] and Rb (right)

A Matlab code describing the charge breeding process is available at the PANDORA collaboration [65], and will be applied to the case of $^7Be^{1+}$ ions: it has been already applied to the capture of Na and Rb ions, giving an excellent agreement with the experimental data (see figure 17). The code is able to track injected particles inside the plasma, and also to store the locations where they are lost, as previously shown in figure 16 (left). More, it is able to give two important informations about the beam-plasma interaction: the density map, showing the spatial distribution of the particles captured by the plasma, figure 16(right), and the energy release map, showing the locations where injected particles loose their energy (see figure 13). The former will help us in finding the necessary plasma dimension to properly trap the injected ions, thus reducing losses; the latter will be used to avoid the possible excitation of plasma instability due to the energy released by the injected particles, as described in [65,67]

## 6 Application as high-intensity X-ray source

Operating in overdense-plasma mode, high-intensity and stable X-ray beams (few keV to 100 keV), will be produced, suitable for non-destructive characterization of samples of interest in the Cultural Heritage field. These beams will be extracted in air (or in a helium atmosphere) and fast tuned in order to promote specific energy domains. In the dedicated X-ray branch we are able to develop specific methods and instrumentations:

i) A modular $\mu$XRD/$\mu$XRF set-up [68,69] for compositional and mineralogical investigation of ancient mineral pigments. The complementary use of X-ray diffraction (XRD) and X-ray fluorescence (XRF) is particularly suited for ancient pigments analysis: XRF gives elemental information on the pictorial layer while XRD allows to identify the mineralogical phases. The continuous component of the X-ray beam emitted by the PANDORA trap will be tuned in order to present the maximum intensity at about 5-6 keV. A pinhole collimator at the exit of the plasma chamber will allow the coupling with X-ray polycapillary lenses. In particular a semi-lens will be used to perform micro-diffraction for mineralogical phases analysis. This device is an optimal choice for the application of the XRD techniques, since it is capable to deliver a quasi-parallel beam to the sample. Diffracted X-rays detection can be performed by a standard SDD detector. The energy selection for performing the diffraction is obtained by applying a digital filter acting on the MCA analyzer. The acquisition of the diffraction pattern is performed by an angular scansion ($\theta - \theta$) of the sample. A contemporary $\mu$XRF analysis will be performed; in this way fluorescence and diffraction data can be combined in order to obtain quantitative information on the pigments (or mixture of pigments). This modular instrument will be used to study the fired pigments in polychrome ceramics belonging to different cultures;

ii) Monochromatic micro X-ray fluorescence [70] devoted to quantitative analysis of metals and provenance studies of archaeological obsidians, pottery and native precious metals. The use of monochromatic radiation gives the opportunity to operate with higher peak-to-background ratio, improving the minimum detection limits (up to the ppb level), thus allowing easier quantitative procedures. The monochromatic $\mu$X-ray fluorescence installed in PANDORA can operate in different energy domains: at low excitation energy (5-6 keV) for non-destructive characterization of light materials and/or surface layers, and at higher beam energy ( 17 keV) for bulk investigation. X-rays generated in the plasma-trap will be collimated in order to get a micro-beam (50-100 microns spot-size); this will be coupled with a doubly curved crystal, positioned in a Rowland circle with the sample. We plan to use two different crystal typologies(e.g. Ge-220 and/or Si-220). Again a modular set-up will be used in order to perform measurements at different incident energies on the same position. The X-rays will be detected by an high performances 20 mm$^2$ SDD detector. The monochromatic $\mu$XRF can be successfully used in non-destructive characterization of ancient metals and in provenance studies of archaeological obsidians, fine ceramics and native precious metals. A peculiar feature of the monochromatic $\mu$XRF instrument is the very high sensitivity, particularly suited for non-destructive trace element analysis of archaeological obsidians and fine pottery.

iii) A X-ray Absorption Near Edge Spectrometry (XANES) instrument [71] for the study of the oxidations state in material of interest in the cultural heritage field. XANES is a well-established technique sensitive to the chemical speciation of the sample constituents. The technique is based on change of X-ray absorption edge induced by the chemical state (i.e. oxidation state) of the material under investigation. The X-ray beam produced by PANDORA can be used to develop a compact XANES instrument operating in a reflection geometry. The use of curved crystals allows to collect X-rays in a large solid angles and focus them to a small spot selecting a narrow energy bandwidth. Crystal and sample are positioned in a peculiar Rowland circle (i.e. Johann geometry). The energy of the beam can be tuned within a definite range (about 100 eV) by properly tilting the crystal optics. The choice of the crystals depends on the chemical specie to be investigated. LiF-200 or Si-111 crystals are appropriate in order to investigate elements like Mn, Fe and Cu elements for applications in cultural heritage field. Finally, the X-ray fluorescence emitted by the sample during the scansion is detected by an high-performance 50-100 mm$^2$ SDD detector. XANES measurements are suitable, for example, for characterization of the oxidation state of chemical elements in degradation processes.

## 7 The PANDORA gear-like methodology

New challenges in different scientific disciplines can be afforded only if a change-of-paradigm approach is argued. Because of its intrinsic nature (i.e. a mixture of electrons and ions at different charge state) plasmas are candidates to perform non-conventional studies in a cross-disciplinary approach. Nuclear decays involving electroweak interaction are affected by free and bound electrons, unveiling an interplay between nuclear and atomic processes, their study has very important astrophysical consequences and can be performed in a controlled plasma, where a collection of different charge states exists in a dynamic equilibrium. At the same time the confined energetic plasma is a unique and valuable light source (a "solar spot") complementing high performance stellar spectroscopy measurements for a better understanding of spectropolarimetric observations in the visible, UV and X-ray domains. In this framework plasma density and temperature must be pushed at the limit of the actual plasma-trap technology. Physical observables like temperature, density and charge state distribution must be directly correlated with in-plasma phenomena in order to collect meaningful information from measurements. Part of the incoming microwave power used for plasma generation is converted to upshifted frequencies as UV and X-rays.

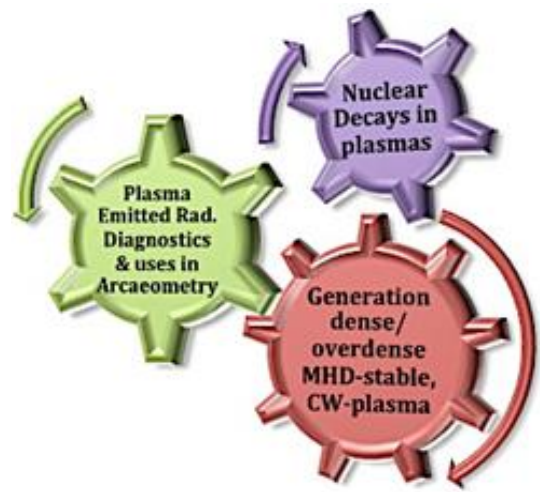

**Fig. 18.** Gear-like diagram sketching the underlying methodology depicted by PANDORA: the strict interplay among plasma science, X-ray diagnostics and applications in material analysis and fundamental studies on nuclear decays rates in the non-neutral matter.

If suitably tuned, the radiation can be used as a probe for materials characterization: different samples can be illuminated by the radiation emitted by the plasma itself. This provides a positive feedback: diagnostics methods use plasma-emitted radiation to improve the power deposition efficiency into the plasma itself (boosting the fluxes of emitted radiation, to be used for elemental analysis purposes), and in turn allows on-line plasma monitoring during nuclear decay measurements. This reciprocal cross-fertilization can be described according to the gear-like diagram of figure 18. The figure highlights the strict relationship among the three disciplines concurring to the satisfaction of the goal of the PANDORA project: the new plasma heating method driven by plasma waves can be exploited provided that new diagnostics are developed, implying over-boosting of plasma density which in turn increases plasma emitted radiation, useful for using the plasma as a source of radiation for Material Characterization. Both dense plasma and its diagnostics are crucial for measuring charge-state and electron-density dependence of the nuclear decays and for providing stellar-like light.